\documentclass[prl,superscriptaddress,twocolumn]{revtex4}

\usepackage{enumerate}
\usepackage{amsfonts,amssymb,amsmath}
\usepackage[]{graphics,graphicx,epsfig}
\usepackage{amsthm}

\begin{document}

\title{Reply to Comment on ``Contextuality in bosonic bunching"}

\author{Pawe\l\ Kurzy\'nski}
\email{cqtpkk@nus.edu.sg}
\affiliation{Centre for Quantum Technologies,
National University of Singapore, 3 Science Drive 2, 117543 Singapore,
Singapore}
\affiliation{Faculty of Physics, Adam Mickiewicz University,
Umultowska 85, 61-614 Pozna\'{n}, Poland}

\author{Akihito Soeda}
\affiliation{Department of Physics, Graduate School of Science, University of Tokyo, 7-3-1 Hongo, Bunkyo-ku, Tokyo, Japan}

\author{Jayne Thompson}
\affiliation{Centre for Quantum Technologies,
National University of Singapore, 3 Science Drive 2, 117543 Singapore,
Singapore}

\author{Dagomir Kaszlikowski}
\email{phykd@nus.edu.sg}
\affiliation{Centre for Quantum Technologies,
National University of Singapore, 3 Science Drive 2, 117543 Singapore,
Singapore}
\affiliation{Department of Physics,
National University of Singapore, 3 Science Drive 2, 117543 Singapore,
Singapore}



\maketitle

The preceding Comment \cite{TA} by Tichy and Andersen is based on the apparent violation of the exclusivity principle by our three-boson system~\cite{us}. It is stated that following our reasoning the three events considered by us are not exclusive which seems to disprove the Letter's conclusion. Moreover, our assumption (ii) of noncontextuality is claimed to be too strong and unreasonable. Finally, a hidden-variable model describing our system is proposed.

Let us recall that the exclusive principle asserts that the sum of probabilities of three pairwise exclusive events can not exceed 1~\cite{Cabello,Fritz}.
Quantum theory is shown to satisfy this principle if the exclusive events can be expressed as pairwise orthogonal projectors.
Although this is certainly a reasonable formulation of exclusive events, the exclusitivity introduced in the Letter is more counterfactual.
Note that our approach is also shared by other researchers such as Yu and Oh of Ref.~\cite{YO}, in which they show that a certain four events are exclusive following their assumptions despite the corresponding projectors being nonorthogonal.

The question of (noncontextual) hidden variables is interesting only for measurements which are argued to be independent under a reasonable hidden-variable model.
In the case of Bell's inequality, a local hidden-variable is chosen to comply with the classical understanding of special relativity, which directly implies that space-like separated measurements do not influence each other.
If one adopts a \textit{nonlocal} hidden-variable model, then even the space-like separated measurements are no longer guaranteed to be independent.

The noncontextuality assumption in the letter was based on the following two observations.
First, the quantum theoretical description of bunching phenomena does not require any interaction between the two bosons and is fully described by a single-particle Hamiltonian.
This motivated us to describe the bunching phenomenon via a model that, just like the quantum description, does not involve interactions.
In addition, what is important for the inequality, is that the particle exchange symmetries do not allow for signalling/disturbance, i.e.\ introducing another photon in the other BS port should not change the marginal scattering probability of the first photon. This is indeed upheld, as each photon still has a 50 \% chance for transmission/reflection. In a sense, the exchange interaction resembles the ``spooky action at a distance" -- it is not a real physical interaction, since it does not allow for information transfer.
Under  such a model we can regard the measurements in the letter as independent.

We argue that the hidden-variable model presented in the Comment \cite{TA} does not prove that their measurements are independent. In fact, the model is clearly contextual.
The comment's model assigns a hidden variable $0 <\lambda_i < 1$ to each boson.
The questions $A_i$ are defined as: will photon ``i" be reflected or transmitted through the beam splitter?
Since for each $A_i$ there are only two possible (exclusive) answers, we can label them as $+1$ and $-1$.
When we measure $A_i$ and $A_j$: if $\lambda_i  > \lambda_j$ the comment's model assigns $A_i = +1$ and $A_j = -1$ (equivalent to {\bf$ \underline{a}_i$ $a_j$} in the letter); and if $\lambda_j  > \lambda_i$ the model assigns $A_i = -1$ and $A_j = +1$  ({\bf$ \underline{a}_j$ $a_i$}).
If we consider pre-assigned $\lambda_i$'s, such that $\lambda_1 < \lambda_2 < \lambda_3$, then measuring $A_2$ and $A_3$ would result in $A_2 = -1$, however measuring $A_2$ and $A_1$ would result in $A_2 = +1$.
Therefore, the outcome of $A_2$ is context dependent. Note, that the above model when applied to the standard contextuality scenario will not only maximally violate the Specker's inequality, but will also maximally violate the Klyachko-Can-Binicioglu-Shumovsky (KCBS) inequality \cite{KCBS} up to its no-disturbance and arithmetic bound of -5, which is also not allowed by the exclusivity principle.

{\it Acknowledgments.}--- This work is funded by the Singapore Ministry of Education
through Academic Research Fund Tier 3 ''Random numbers
from quantum processes" Grant No. MOE2012-T3-1-009 and
by the Singapore National Research Foundation. P. K. and
D. K. are also supported by the Foundational Questions
Institute (FQXi). The authors acknowledge correspondence with Malte Tichy and Christian Andersen.



\begin{thebibliography}{99}

\bibitem{TA}
M. C. Tichy and C. K. Andersen, Phys. Rev. Lett. {\bf 113}, 138901 (2014).

\bibitem{us}
P. Kurzy\'nski, A. Soeda, J.  Thompson and D. Kaszlikowski, Phys. Rev. Lett. {\bf 112}, 020403 (2014)

\bibitem{Cabello}
A. Cabello, Phys. Rev. Lett. {\bf 110}, 060402 (2013)

\bibitem{Fritz}
T. Fritz,	A. B. Sainz,	R. Augusiak, J. Bohr Brask, R. Chaves, A. Leverrier, and A. Acin, Nat. Comm.  {\bf 4}, 2263 (2013)

\bibitem{YO}
S. Yu, C. H. Oh, Phys. Rev. Lett. {\bf 108}, 030402 (2012)






\bibitem{KCBS}
A. A. Klyachko, M. A. Can, S. Binicioglu, and A. S. Shumovsky,
Phys. Rev. Lett. {\bf 101}, 020403 (2008).


\end{thebibliography}
\end{document}